\documentclass[useAMS,usenatbib]{mn2e}

\usepackage{graphicx}
\usepackage{bm}
\usepackage{natbib}
\usepackage{subfig}
\usepackage{subfloat}

\newcommand{\be}{\begin{equation}}
\newcommand{\ee}{\end{equation}}

\newcommand{\ba}{\begin{eqnarray}}
\newcommand{\ea}{\end{eqnarray}}

\title[Hall Effect in Neutron Star Crusts]{Hall Effect in Neutron Star Crusts: Evolution, Endpoint and Dependence on Initial Conditions}

\author[K.N.~Gourgouliatos \& A. Cumming]{\parbox{\textwidth}{K.N.~Gourgouliatos\thanks{E-mail: kostasg@physics.mcgill.ca}$^{1,2}$, A.~Cumming$^{1}$
}\vspace{0.4cm}\\
\parbox{\textwidth}{ $^{1}$ Department of Physics, McGill University, 3600 rue University, Montr\'eal, Qu\'ebec H3A 2T8, Canada \\
 $^{2}$ Centre de Recherche en Astrophysique du Qu\'ebec Fellow\\ } }

\begin{document}

\date{Accepted -. Received -; in original form -}
\pagerange{\pageref{firstpage}--\pageref{lastpage}} \pubyear{-}
\maketitle

\label{firstpage}

\begin{abstract}
We present new simulations of the evolution of axially symmetric magnetic fields in neutron star crusts under the influence of the Hall effect and subdominant Ohmic dissipation. In the Hall effect, differential rotation of the electron fluid generates toroidal field by winding of the poloidal field. For this reason, we focus on the influence of the initial choice of the electron angular velocity profile on the subsequent and long term magnetic evolution. Whereas previous simulations have generally chosen angular velocities increasing outwards, corresponding to the lowest order Ohmic mode in the crust, a more realistic choice is an angular velocity decreasing outwards, corresponding to the MHD equilibrium field that is likely present at the time of crust formation. 
We find that the evolution passes through three basic phases. The early evolution is a response to the initial conditions. During the second phase the field consists of poloidal and toroidal components which eventually relax to an isorotation state in which the angular velocity of the electrons becomes constant along poloidal magnetic field lines, causing Hall evolution to saturate. In the third phase the field dissipates slowly while maintaining isorotation. We discuss the implications for the long term field structure and observable properties of isolated neutron stars.
\end{abstract} 


\section{Introduction}
Neutron stars are known to host magnetic fields reaching $10^{15}$G in the case of magnetars \citep{Nakagawa:2009}, while some millisecond pulsars have fields down to $10^{8}$G \citep{Possenti:2003}. This broad range of magnetic field intensities combined with the fact that magnetars show activity powered by the magnetic field \citep{Thompson:1995} and usually have relatively young inferred ages provides evidence that the magnetic field evolves with time. 

The evolution of the magnetic field inside the neutron star is mediated by three physical processes: Hall drift, which is the advection of the magnetic field because of the motion of the free electrons; Ohmic dissipation which converts magnetic energy to heat because of the finite conductivity of the crust; and ambipolar diffusion which is the interaction of the electric currents with the neutrons deeper inside the neutron star \citep{Jones:1988, Goldreich:1992}. Depending on the details of the magnetic field intensity, temperature, density, lattice impurities and electron fraction, each process is dominant in different regions of the neutron star. For magnetic fields of the order of $10^{13}$G and higher, the Hall timescale is at least an order of magnitude shorter than the Ohmic time scale in the crust and the evolution will be mainly due to the Hall effect, while, as long as the magnetic field is weaker than $\sim 10^{15}$G any Lorentz force will be balanced by elastic forces in the crust. Thus, Ohmic dissipation will be important in current sheets and ambipolar diffusion shall take place in the core of the neutron star. 

The evolution of the magnetic field due to the Hall effect is non-linear and there are only a few analytical solutions known, i.e.~non-linear whistler waves propagating along a uniform field \citep{Goldreich:1992}, thus a simulation is essential for more complicated structures. Many authors have performed simulations of the Hall effect to investigate its influence on the global evolution of the magnetic field. \cite{Urpin:1991} simulated the evolution of a toroidal field in a constant density sphere and later included a poloidal field in the evolutionary scheme, concluding that the field generates higher multipole modes that oscillate \citep{Shalybkov:1997}. \cite{Naito:1994} studied the evolution of the magnetic field in the opposite limit, where the Hall effect is subdominant compared to the Ohmic dissipation showing that the Hall effect excites higher order modes. \cite{Hollerbach:2002, Hollerbach:2004} studied the evolution of the magnetic field under the Hall effect in a uniform and a stratified crust using a spectral code, finding the development of fine structures that make the dissipation of the magnetic field faster. \cite{Pons:2007} developed a mixed spectral and finite difference code and calculated the evolution of the magnetic field using realistic choices of the microphysical parameters, finding that in such conditions the field tends to move closer to the inner crust. \cite{Vigano:2012} developed a staggered finite difference code which was later combined with cooling codes to unify the observational properties of neutron stars \citep{Vigano:2013}. From a different perspective, \cite{Geppert:2013} have utilised the Hall effect to explain the formation of magnetic spots that can be associated with pulsar radio emission.  

The results of these different numerical simulations agree in that, while the Hall effect leads to transfer of energy between toroidal and poloidal components and generation of higher order structure in the poloidal field, it does not appear to result in a rapid dissipation of the field, instead it saturates. Rapid dissipation had been suggested by analytic works on the Hall effect. \cite{Goldreich:1992} proposed that the non-linear Hall term would drive a cascade to small scales where Ohmic dissipation would occur. \cite{Vainshtein:2000} approached the problem of the evolution of a plane parallel magnetic field in a stratified density medium and using both analytical and numerical arguments found that the field obeys a Burgers-type equation which leads naturally to the formation of shocks hosting current sheets in which magnetic energy can be dissipated on a Hall time. \cite{Reisenegger:2007} studied analytically the evolution of an axially symmetric field, showing that a purely toroidal field obeys a Burger's type differential equation similarly to the plane parallel case, while a weak poloidal field superimposed to a toroidal field will lead to a long-wavelength instability as the one discussed by \cite{Rheinhardt:2002} in a plane-parallel geometry.

Simulations in cartesian boxes do find that the Hall effect leads to a turbulent cascade, and the scaling and anisotropy of the turbulent spectrum has been well-studied \citep{Biskamp:1996, Wareing:2009, Cho:2009}. \cite{Wareing:2009b, Wareing:2010} found that although turbulence in evident in the power spectrum, the temporal evolution is different from typical MHD turbulence, with the formation of long-lasting structures.  The reason for the different behaviour between crust and cartesian box simulations is not clear. Neutron star simulations have so far been axisymmetric, which perhaps suppresses some modes of decay. Alternatively, Hall cascade simulations are carried out at constant density and do not include the shock formation that results from an electron density gradient.    

On the other hand, stationary states under the Hall effect have been found for neutron stars. \cite{Cumming:2004} found a stationary state for a purely poloidal field of dipole structure supported by a uniformly rotating electron fluid. Rigidly rotating electrons do not wind the poloidal field into toroidal field, allowing for the existence of a steady state. \cite{Gourgouliatos:2013, Gourgouliatos:2013b} generalised this result, finding stationary solutions corresponding to Hall equilibria with mixed toroidal and poloidal fields, where all the toroidal field is concentrated in poloidal loops that close inside the star. They showed that Hall equilibria with external dipole fields have poloidal current distributions that correspond to rigid rotation of the electron fluid. 

 \cite{Gourgouliatos:2013} pointed out that the relevance of these Hall equilibria states is not clear for neutron star magnetic field evolution since the Hall effect by itself does not necessarily lead to an equilibrium state. Hall evolution conserves magnetic energy, and does not have a preferred state. We address this major unanswered question here and study the long term evolution of the field under the joint action of the Hall effect and Ohmic dissipation. We find that the field structure does adjust so that it lies very close to a Hall equilibrium state. Rather than rigid rotation of the electrons, the equilibrium state is one of isorotation in which different field lines rotate with different angular velocities, but with rigid rotation along each individual field line, analogous to Ferraro's law in MHD \citep{Ferraro:1937}. 

We also address the sensitivity of Hall evolution to the choice of initial conditions. Independent of the Hall effect studies, significant work has been done on the Magnetohydrodynamic (MHD) equilibria of stars addressing both the question of equilibrium and stability in normal stars and neutron stars \citep{Braithwaite:2006, Haskell:2008, Lander:2009, Aly:2012, Glampedakis:2012, Lander:2012, Lander:2013}. These studies suggest that a toroidal field containing comparable energy with the poloidal field is essential for stability, with the toroidal field concentrated in the equatorial zone. Such a field is likely to be present when the crust forms in a young neutron star, and so would represent the initial condition for evolution under the Hall effect. However, Hall simulations have mostly used Ohmic eigenmodes as initial conditions, sometimes containing a broadly distributed toroidal field. We investigate MHD equilibria as initial states for Hall evolution here and show that the early evolution is qualitatively different from the Ohmic decay eigenmode initial condition because the differential rotation in the electron fluid is of opposite sign.

An outline of the paper is as follows. We give the basic equations that describe the Hall effect in Section 2. We study toroidal fields in Section 3, which are used as tests for our simulation as they can be understood analytically. We study mixed poloidal and toroidal fields in Section 4, where we show that the early evolution is sensitive to the initial profile of electron angular velocity, and find that the long term evolution of the field is a state of isorotation, in which the electron angular velocity is constant along poloidal field lines. We discuss these results and their observational implications in Section 5. We conclude in Section 6. The details of our numerical method and initial conitions for the simulations are presented in the appendix.

\section{Mathematical Formulation and Crust Model}
\label{FORMULATION}

\subsection{Mathematical Formulation}

We express the magnetic field in poloidal and toroidal components and trace their evolution in terms of two scalar functions $\Psi$ and $I$. The magnetic field is 
\begin{eqnarray}
\bm{B}=\nabla \Psi \times \nabla \phi + I \nabla \phi\,,
\label{B}
\end{eqnarray}
where we use spherical coordinates $(r, \theta, \phi)$, and $\nabla\phi = \hat{\phi}/r\sin\theta$, with $\hat{\phi}$ the unit vector in the $\phi$ direction. $2\pi \Psi$ is the poloidal magnetic flux passing through a spherical cap of radius $r$ and opening angle $\theta$, and $cI/2$ is the poloidal electric current passing through the same spherical cap, where $c$ is the speed of light. The crust of the neutron star can be approximated by a crystal lattice where only electrons have the freedom to move, thus the electric current is $\bm{j}=-n_{\rm e}{\rm e} \bm{v}$, where $n_{\rm e}$ is the electron density, ${\rm e}$ is the elementary electric charge and $\bm{v}$ is the electron velocity. For a finite electrical conductivity $\sigma$, the electric field is $\bm{E}=-\bm{v}\times \bm{B}/c +\bm{j}/{\sigma}$. Using Amp\`ere's law $\bm{j}=(c/4\pi)\nabla\times\bm{B}$, the induction equation is then
\begin{eqnarray}
\frac{\partial \bm{B}}{\partial t} = -\frac{c}{4 \pi{\rm e}}\nabla \times  \left(\frac{\nabla \times \bm{B}}{n_{\rm e}} \times \bm{B}\right) -\frac{c^{2}}{4\pi} \nabla \times \left(\frac{\nabla \times \bm{B}}{\sigma}\right)\,.
\label{dB}
\end{eqnarray}
The first term on the right hand side of equation (\ref{dB}) is the Hall evolution term and the second one gives rise to Ohmic dissipation.

We rewrite the induction equation in terms of the scalar functions $\Psi$ and $I$.
Following \cite{Reisenegger:2007}, we define the quantity 
\begin{equation}
\chi={c\over 4\pi {\rm e}n_{\rm e} r^{2}\sin^{2}\theta}\,,
\end{equation}
and also write the toroidal current 
\begin{equation}
\bm{j_T}={c\over 4\pi}\nabla\times\bm{B_P}=-{c\over 4 \pi} \Delta^{*}\Psi \nabla \phi\,,
\end{equation}
in terms of the angular velocity of the electrons,
\begin{eqnarray}
\Omega= -{j_T\over n_{\rm e} {\rm e} r\sin\theta}=\chi \Delta^{*} \Psi\,,
\label{OMEGA}
\end{eqnarray}
where the Grad-Shafranov operator is
\begin{equation}
\Delta^{*}=\frac{\partial^{2}}{\partial r^{2}} +\frac{\sin\theta}{r^{2}}\frac{\partial}{\partial \theta}\left(\frac{1}{\sin\theta}\frac{\partial}{\partial \theta}\right).
\end{equation}
Using the decomposition of the field in toroidal and poloidal components we find that the evolution of the scalar functions is
\begin{eqnarray}
\frac{\partial \Psi}{\partial t} +r^{2}\sin^{2}\theta\chi (\nabla I \times \nabla \phi)\cdot \nabla \Psi =\frac{c^{2}}{4 \pi \sigma}\Delta^{*}\Psi\,,
\label{dPSI}
\end{eqnarray}
\begin{eqnarray}
\frac{\partial I}{\partial t} +r^{2}\sin^{2}\theta (\left( \nabla \Omega \times \nabla \phi \right) \cdot \nabla \Psi + I \left(\nabla \chi \times \nabla \phi \right) \cdot \nabla I ) \nonumber\\= \frac{c^{2}}{4 \pi \sigma}\left(\Delta^{*}I+\frac{1}{\sigma}\nabla I\times \nabla \sigma\right) \,.
\label{dI}
\end{eqnarray}

The evolution of the poloidal field can be understood through equation (\ref{dPSI}). The left hand side is the Lagrangian derivative of $\Psi$ with a velocity field $\bm{w}_{p}=r^{2}\sin^{2}\theta\chi (\nabla I \times \nabla \phi)$. Thus the poloidal field evolves by advection along surfaces of constant $I$, so that essentially the evolution of the poloidal field is mediated by the toroidal field. According to equation (\ref{dI}) the surfaces of constant $I$ change with time, and there are terms depending on $\Psi$, thus there is a feedback loop. The right-hand-side of equation (\ref{dPSI}) describes the Ohmic losses in the poloidal field because of finite conductivity. 

The Hall part of the evolution of the toroidal field (Equation \ref{dI}) has two terms. The first term in the bracket in the left-hand-side of equation (\ref{dI}) describes the winding of the poloidal field into toroidal when $\Omega$ is not constant along a poloidal field line, while in the case of isorotation this term vanishes. The second term inside the bracket is the advection of $I$ along surfaces of constant $\chi$, this term depends on the density and the geometry of the system. The terms on the right-hand-side give the Ohmic dissipation of the toroidal field. 

We have developed an explicit finite-difference code to solve equations (\ref{dPSI}) and (\ref{dI}). The details of the code are given in the Appendix, as well as the initial conditions for all of the numerical runs shown later in the paper.

\subsection{Crust model}

The Ohmic and Hall terms vary substantially across the crust because the electron density $n_{\rm e}$ and electrical conductivity $\sigma$ change by large factors. We include a factor of 100 variation in $n_{\rm e}$ across the crust, which we take to be of thickness $0.1 r_*$ where $r_*=10\ {\rm km}$ is the neutron star radius. The baryon number density at the base of the crust is $n_b\approx 0.08\ {\rm fm^{-3}}$, and the electron fraction is $Y_{\rm e}\approx 0.03$, giving $n_{\rm e}=Y_{\rm e}n_b\approx 2.5\times 10^{36}\ {\rm cm^{-3}}$, \citep{Chamel:2008}. We take this as our base density, and following \cite{Cumming:2004}, we use the scaling $n_{\rm e}\propto z^4$ where $z$ is the depth into the crust from the surface. Specifically, we choose $n_{\rm e}\propto[(1.0463 r_{*}-r)/0.0463r_{*}]^4$ which drops by a factor of 100 over the crust. Note that because $Y_{\rm e}$ increases outwards by an order of magnitude across the crust, the range of mass density that we model is from $\rho\approx 10^{14}\ {\rm g\ cm^{-3}}$ at the crust-core boundary to $\rho\approx 10^{11}\ {\rm g\ cm^{-3}}$ above neutron drip in the outer crust. The solid part of the outer layers extends to $\rho<10^{11}\ {\rm g\ cm^{-3}}$, depending on temperature. This lower density region is not included in our code to keep the timestep manageable. However, we note that for a $10^{14}\ {\rm G}$ field, the density where the magnetic energy density is comparable to the shear modulus is close to $10^{11}\ {\rm g\ cm^{-3}}$. This means that at low density, the field can overcome the material strength of the crust as it evolves, and so replacing these outer layers by a vacuum boundary condition may be a reasonable approximation.

With the electron density profile set, we then take $\sigma\propto n_{\rm e}^{2/3}$, which is somewhere between the density scalings expected for $\sigma$ based on phonon scattering or impurity scattering \citep{Cumming:2004}. The conductivity ranges from $1.8\times 10^{23} {\rm s^{-1}}$ at the top of the crust to $3.6\times 10^{24}\ {\rm s^{-1}}$ at the base, which are appropriate values for either phonon scattering in a crust with $T\approx 2\times 10^8\ {\rm K}$, or impurity scattering with impurity parameter $Q_{\rm imp}\approx 3$. The values of conductivity are of the same order, but a smaller range, than those used by \cite{Vigano:2012} (they used $\sigma=2.2\times 10^{22}$--$3.1\times 10^{25}\ {\rm s^{-1}}$). Note that we do not include the thermal evolution of the crust and its feedback on the conductivity in our simulations. Instead, our focus is on the physics of the Hall effect and so it is a good starting point to adopt this non-evolving conductivity profile. The Ohmic timescale for our crust model is 
\begin{equation}
t_{\rm Ohm} \sim {4\pi \sigma L^2\over c^2} = 13.5\ {\rm Myr}\ \left({L\over 1\ {\rm km}}\right)^2\left({\sigma\over 3\times 10^{24}\ {\rm s^{-1}}}\right),
\end{equation}
and the Hall timescale is
\begin{equation}
t_{\rm Hall} \sim {4\pi {\rm e} L^{2} n_{\rm e} \over cB} = {1.6\ {\rm Myr}\over B_{14}}\ \left({L\over 1\ {\rm km}}\right)^2\left({n_{\rm e}\over 2.5\times 10^{36}\ {\rm cm^{-3}}}\right),
\end{equation}
where we insert the full crust thickness and the density and conductivity at the base of the crust. Lower density regions of the crust have a significantly shorter Hall timescale, as the density, field strength, and lengthscales change across the crust. The Hall timescale also becomes longer with time as the magnetic field dissipates. Throughout the paper when we refer to the Hall timescale we shall refer to the average initial value which is in the range of $100$--$200$ kyrs.    

The ratio of the Hall and Ohmic terms can be measured with the magnetic Reynolds number $R_{M}=\sigma |B|/(n_{\rm e} {\rm e} c)$. Although $\sigma$, $n_{\rm e}$ and $B$ vary strongly inside the crust, the change of $R_{M}$ is less dramatic, as in realistic crusts all the three quantities decrease moving closer to the surface. In our simulations we choose initial conditions with a maximum $R_{M}$ in the range of $50-100$.  The detailed choices of the fields are discussed in the Appendix, in general they correspond to surface magnetic fields in the range of $5\times 10^{13}-10^{14}$G.  The Reynolds number $R_M$ is the important parameter for our numerical models. Our results are applicable for different choices of $\sigma$ and $B$ as long as they are chosen so that $\sigma B$ is approximately constant so that the Reynolds number is the same, and the Hall and Ohmic timescales are scaled appropriately. For example, our results apply to a $B=10^{15}\ {\rm G}$ magnetar with a conductivity that is ten times smaller than assumed here (for example with an impurity parameter of $\approx 30$ as suggested for the inner crust by \citealt{Vigano:2013}). However, in that case, the Ohmic and Hall timescales are both smaller by a factor of ten, so the results shown later in the paper should have their times divided by a factor of ten.

\section{Pure toroidal fields}
\label{TOROIDAL}

The evolution of toroidal fields has been studied by many authors, both analytically \citep{Vainshtein:2000,Reisenegger:2007} and numerically, most recently by \cite{Kojima:2012} and \cite{Vigano:2012}. We have run several cases of toroidal fields only as they are a useful check on our numerical code. Here, we would like to point out a qualitiative difference in the evolution of toroidal fields under the Hall effect depending on the choice of density profile in the crust.

A pure toroidal field does not generate any poloidal field because of the Hall effect and Ohmic dissipation. $I$ is advected with velocity $\bm{w_{T}}= r^{2}\sin^{2}\theta I \nabla \chi \times \nabla \phi$, in other words along the constant $\chi$ surfaces \citep{Reisenegger:2007}. In a constant density configuration, $\chi$ depends only on cylindrical radius $R=r\sin\theta$, and so this velocity field is $\bm{w_{T}}=-(cI/2 \pi {\rm e} n_{\rm e} R^{2})\hat{\bm{z}}$, where $(R, \phi, z)$ are the cylindrical polar coordinates. There is motion in the $z$ direction only, and the evolution of the magnetic field is driven by the geometric term $R^2$ in $\chi$. A positive toroidal field will move in the $-\bm{\hat{z}}$ direction, while a toroidal field of the opposite polarity will move in the $+\bm{\hat{z}}$ direction. The field will be pushed either to the crust-core boundary, or to the surface of the star. In either case, the fact that the velocity is proportional to $I$ leads to steepening of the profile of $I$ \citep{Vainshtein:2000}, and the field will form a shock where Ohmic dissipation proceeds in a Hall time scale.

This evolution changes drastically if we include a realistic electron density for the neutron star crust assuming a spherically symmetric profile $n_{\rm e}=n_{\rm e}(r)$. The advection velocity becomes
\begin{eqnarray}
\bm{w_{T}}=\frac{cI}{4\pi {\rm e} n_{\rm e} R}\left[\cos\theta \frac{n_{\rm e}'}{n_{\rm e}} \hat{\bm{R}} -\left(\frac{2}{R} +\sin\theta \frac{n_{\rm e}'}{n_{\rm e} }\right)\hat{\bm{z}}\right] \,,
\end{eqnarray} 
where prime denotes derivation with respect to $r$. This velocity profile leads to motion both in $z$ and $R$, and if the variation in $n_{\rm e}$ is much faster than $1/R^2$ (which is indeed the case for a neutron star crust), the surfaces of constant $\chi$ are almost spherical and so the motion is along almost spherical surfaces, except for very close to the poles. This leads to a behaviour qualitatively opposite to the constant density case: a positive toroidal field will travel along the spherical $\chi$ surfaces to the north,  a negative toroidal field moves towards the south. The surfaces of constant $\chi$ intersect with the crust-core boundary, so the field will be pushed eventually there, but unlike the constant density case they do not intersect with the surface of the star, therefore a purely toroidal field will never be pushed to the surface of the neutron star, as was indeed seen by \cite{Hollerbach:2004}.

We remark the similarity of evolution of a toroidal field to a plane parallel one $\bm{B}=B_{y}(x, z)\bm{\hat{y}}$, where the density has been rescaled by $\tilde{n}_{e}(x,z) =x^{2}n_{e}(x,z)$. This is because in axial symmetry and constant electron density the available electric charges to support a loop of toroidal field are proportional to the radius, while for the same enclosed electric current the magnetic field depends on the radius by $B_{\phi}\propto \frac{I}{R}$. Thus the evolution of an axially symmetric toroidal field and a plane parallel one are similar if in the latter the electron density profile is proportional to the square of the  distance from a plane of reference.

We have confirmed the analytical considerations discussed above through simulation. We have used a variety of initial conditions, including multipoles with $\ell=1$ and $\ell=2$ of various polarities, leading to behaviours which are in accordance with the analytical considerations discussed above: the field evolves with $I$ being advected on surfaces of constant $\chi$.  As expected, a constant density profile gives qualitative opposite evolution, and eventually counter-intuitive results, if they are applied in realistic crusts. Energy dissipation is indeed driven by the Hall effect: energy is dissipated at the shocks on a Hall timescale. This continues until the magnetic field weakens so that the Ohmic timescale is similar to the Hall ($R_M\sim 1$), and thereafter the dissipation proceeds on the Ohmic timescale.

\section{Evolution of mixed fields}
\label{POLOIDAL}

Below we examine the evolution of mixed poloidal and toroidal fields. If the initial state contains only a poloidal field, it will generate a toroidal field, in contrast to the evolution of a pure toroidal field which remains toroidal. Even in the exceptional case where the initial state is a pure poloidal field in Hall equilibrium, for example with rigidly-rotating electrons, Ohmic decay will push the system out of Hall equilibrium leading to the generation of a toroidal field and subsequent evolution. Therefore, the presence of a poloidal field leads to mixed toroidal-poloidal field. 

In this section we simulate such cases for different initial electron angular velocity profiles.  The magnetic dipole moment used in the initial states is always chosen to point in the $+\bm{\hat{z}}$ direction.  
The detailed initial profiles of $\Psi$ and $I$ for all simulations are given in the Appendix. 

\begin{figure}
\centering
\includegraphics[width=1.15\columnwidth]{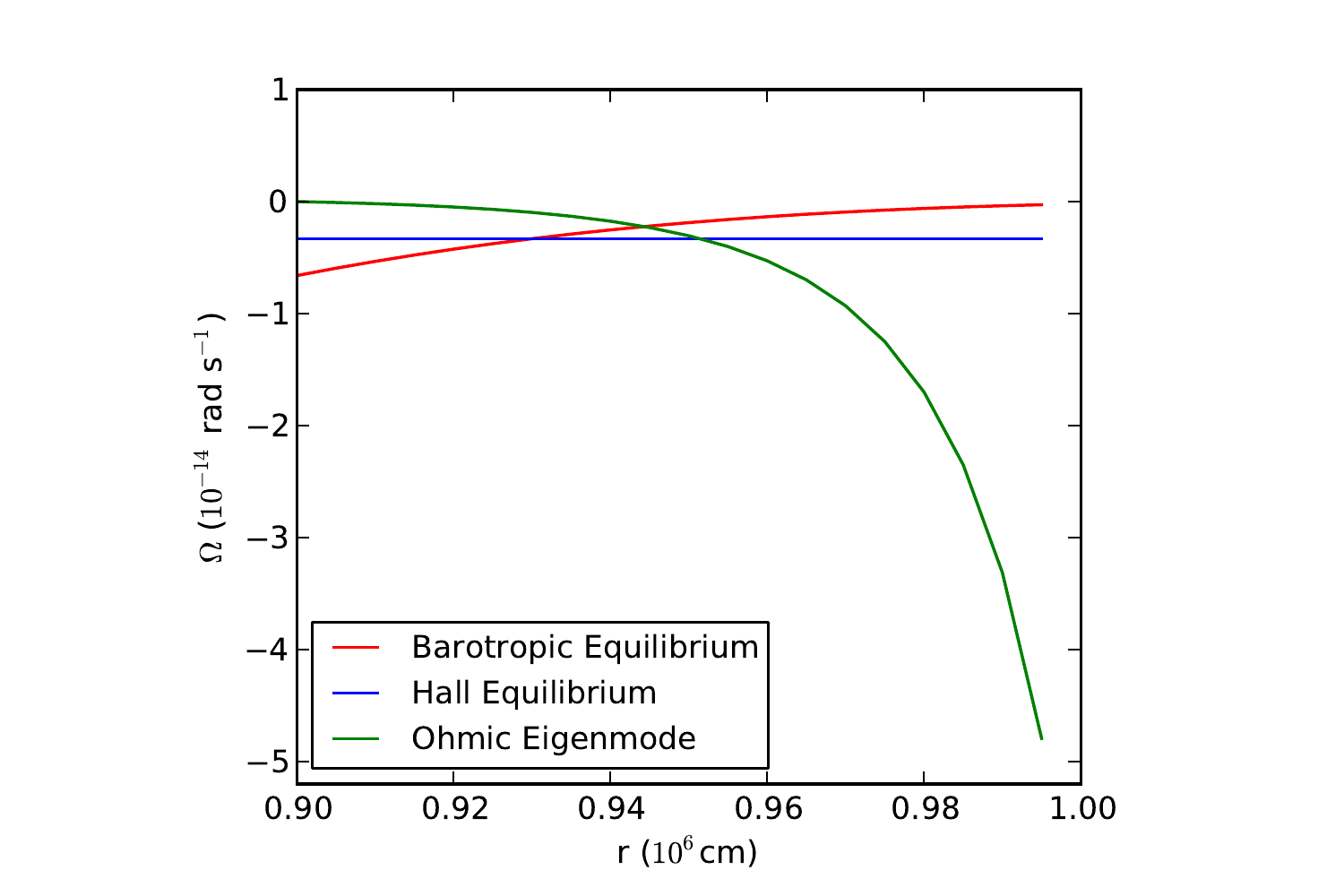}
\caption{$\Omega(r)$ for the initial conditions chosen. The Hall equilibrium we used has a constant $\Omega$ which is shown in blue. The barotropic equilibrium, shown in red, has the faster electrons inwards because $Y_{\rm e}$ increases with $r$. The Ohmic eigenmode is shown in green.}
\label{Fig:1}
\end{figure}

\subsection{Purely poloidal initial field}
\label{PPOLOIDAL}

\subsubsection{The choice of initial conditions and electron angular velocity profile}

Let us assume that the initial field is only poloidal. The current is toroidal, and the electron angular velocity is given in terms of $\Psi$ by equation (\ref{OMEGA}). For example, a dipole field $\Psi=\sin^{2}\theta ~g(r)$ is related to $\Omega(r)$ by $g''-2g/r^{2}=(4 \pi {\rm e} n_{\rm e}/c)r^{2}\Omega(r)$. Assuming that the initial field is a poloidal dipole it will initially generate a toroidal quadrupole field through $\frac{\partial B_{\phi}}{\partial t}=r\sin\theta B_{r}\frac{\partial \Omega}{\partial r}$. \cite{Cumming:2004} demonstrated that a poloidal dipole field is in a neutrally stable, Hall equilibrium if the electron fluid rotates uniformly. For an initial dipole field without uniform rotation $\Omega \neq$const.~a quadrupolar toroidal field shall appear, which can be of either polarity, which is determined by the product of $B_{r}\frac{\partial \Omega}{\partial r}$. 

Therefore the choice of the initial profile $\Omega(r)$ is of crucial importance. Depending on the polarity of the quadrupolar field and its intensity, the magnetic field shall follow different evolutionary paths during its early evolution. Previous studies have chosen a field corresponding to the fundamental Ohmic eigenmode for constant $\sigma$, satisfying the equation $\nabla \times (\nabla \times \bm{B}) =\alpha \bm{B}$, with $\alpha$ being a constant  \citep{Hollerbach:2002, Vigano:2012} or a similar profile \citep{Kojima:2012}. The profile of electron angular velocity for the $\ell=1$ lowest radial order Ohmic mode is shown as the green curve in Figure \ref{Fig:1}: the angular velocity increases sharply outwards. We can understand this from the equation for the Ohmic eigenmodes, $\Delta^*\Psi \propto \sigma\Psi$, which gives 
\begin{equation}
\Omega_{\rm Ohmic}=\chi \Delta^*\Psi \propto \chi\sigma\Psi\propto {\sigma\over n_{\rm e}}.
\end{equation}
For the constant conductivity case, $\Omega$ therefore increases strongly outwards as $n_{\rm e}$ decreases. Even for a real crust in which the variation of $\sigma$ with $r$ is taken into account, we expect that the Ohmic modes will still have $\Omega$ increasing outwards, because the scaling of $\sigma$ with $n_{\rm e}$ is generally less than $\sigma\propto n_{\rm e}$ \citep{Cumming:2004}.

If the Hall timescale is short compared to the Ohmic time, the field will not have had time to relax to its lowest order Ohmic mode, which makes this an unlikely initial condition for the field. A more realistic magnetic field configuration at the time when the crust forms is an MHD equilibrium \citep{Gourgouliatos:2013}, because crust formation is delayed until the star cools below the melting temperature which takes many Alfv\'en crossing times. 
The detailed structure is still an open question, but a reasonable starting point is a barotropic equilibrium. Such equilibria are sensitive to the mass density of the neutron star. 
A dipole poloidal field being in barotropic MHD equilibrium obeys the equation $\Delta^{*}\Psi +S_{B}r^{2}\sin^{2}\theta \rho=0$, where $S_{B}$ is an appropriate normalization constant, $\rho$ is the mass density, related to the electron number density by $n_{\rm e}= Y_{\rm e} \rho$, where $Y_{\rm e}$ is the electron number per unit mass. Equation (\ref{OMEGA}) then gives
\begin{equation}
\Omega_{\rm MHD}=\frac{S_{B} c}{4 \pi \rm{e} Y_{\rm e}}\propto {1\over Y_{\rm e}},
\end{equation}
which decreases outwards as $Y_{\rm e}$ increases, opposite to the Ohmic mode. The electron angular velocity is shown as the red curve in Figure \ref{Fig:1}.

The intermediate case between $\Omega$ increasing outwards and decreasing outwards is obviously a rigidly-rotating state with $\Omega$ independent of $r$, or in other words a Hall equilibrium (blue curve in Fig.~\ref{Fig:1}). We have simulated all three cases shown in Figure \ref{Fig:1}. The difference in the direction of the differential rotation leads to a qualitatively different evolution at early times, as we will describe. Note that we assume a dipole structure for the initial field, so there is no angular dependence of $\Omega$ initially. To allow comparison we have chosen the total magnetic energy in the system to be the same in all three cases, which leads to slightly different surface fields. 
\begin{figure*}
\centering
\includegraphics[width=1.60\columnwidth]{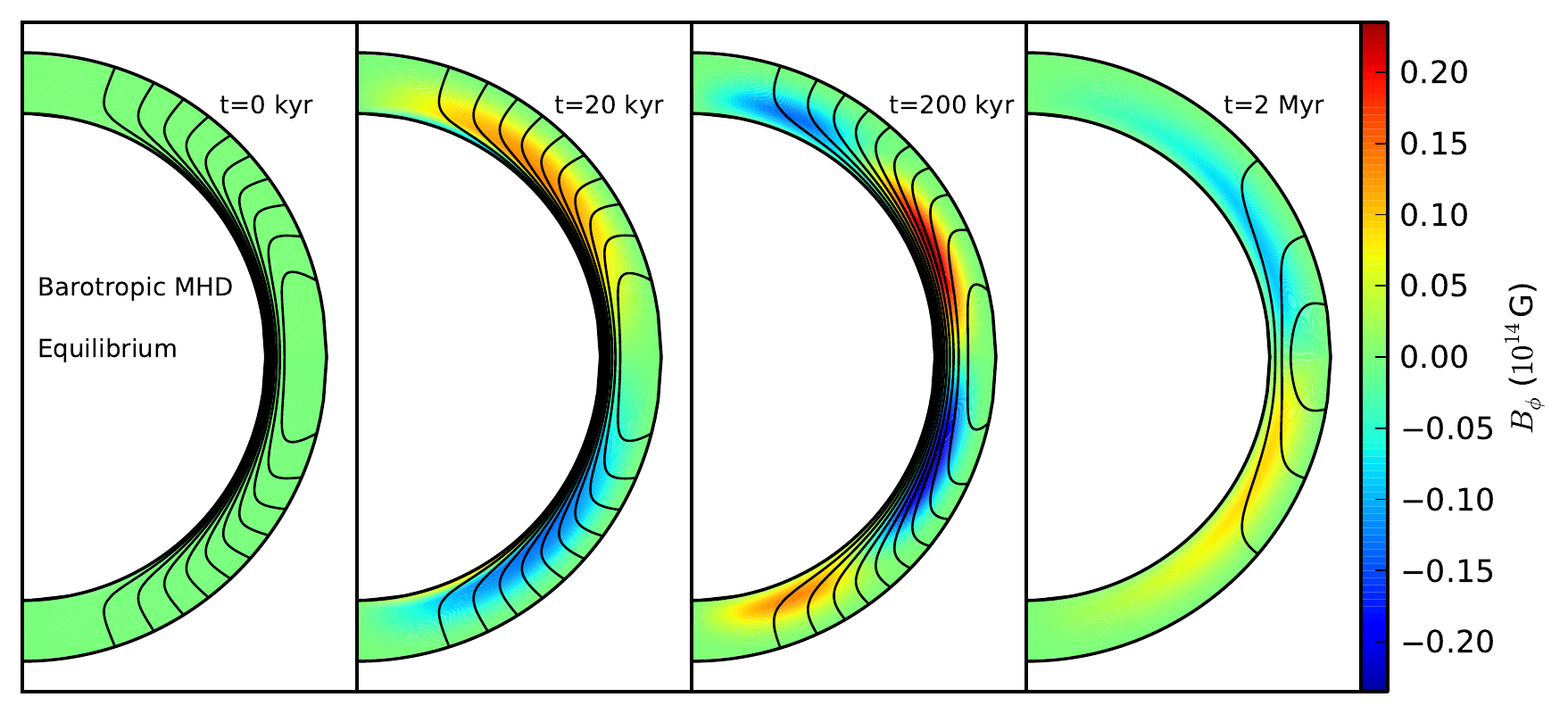}
\includegraphics[width=1.60\columnwidth]{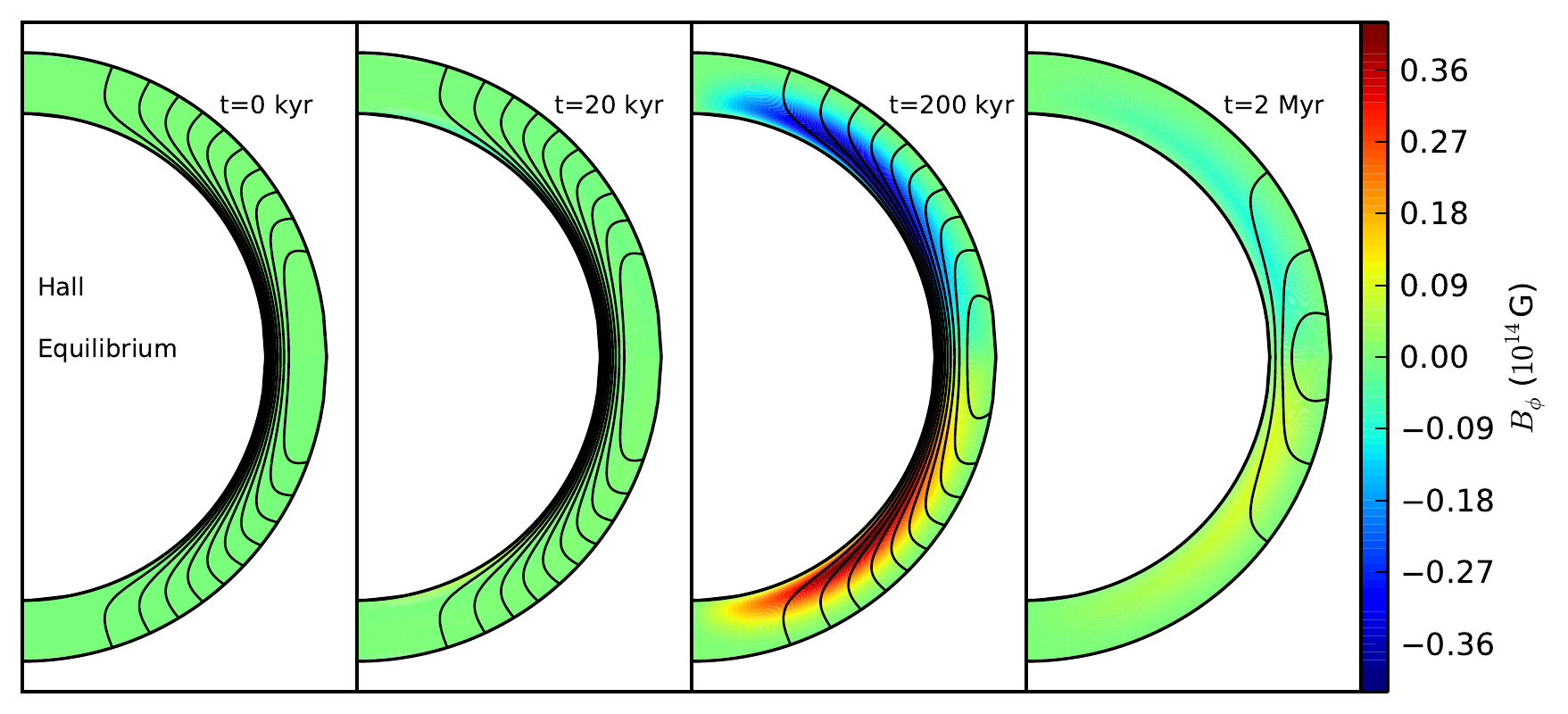}
\includegraphics[width=1.60\columnwidth]{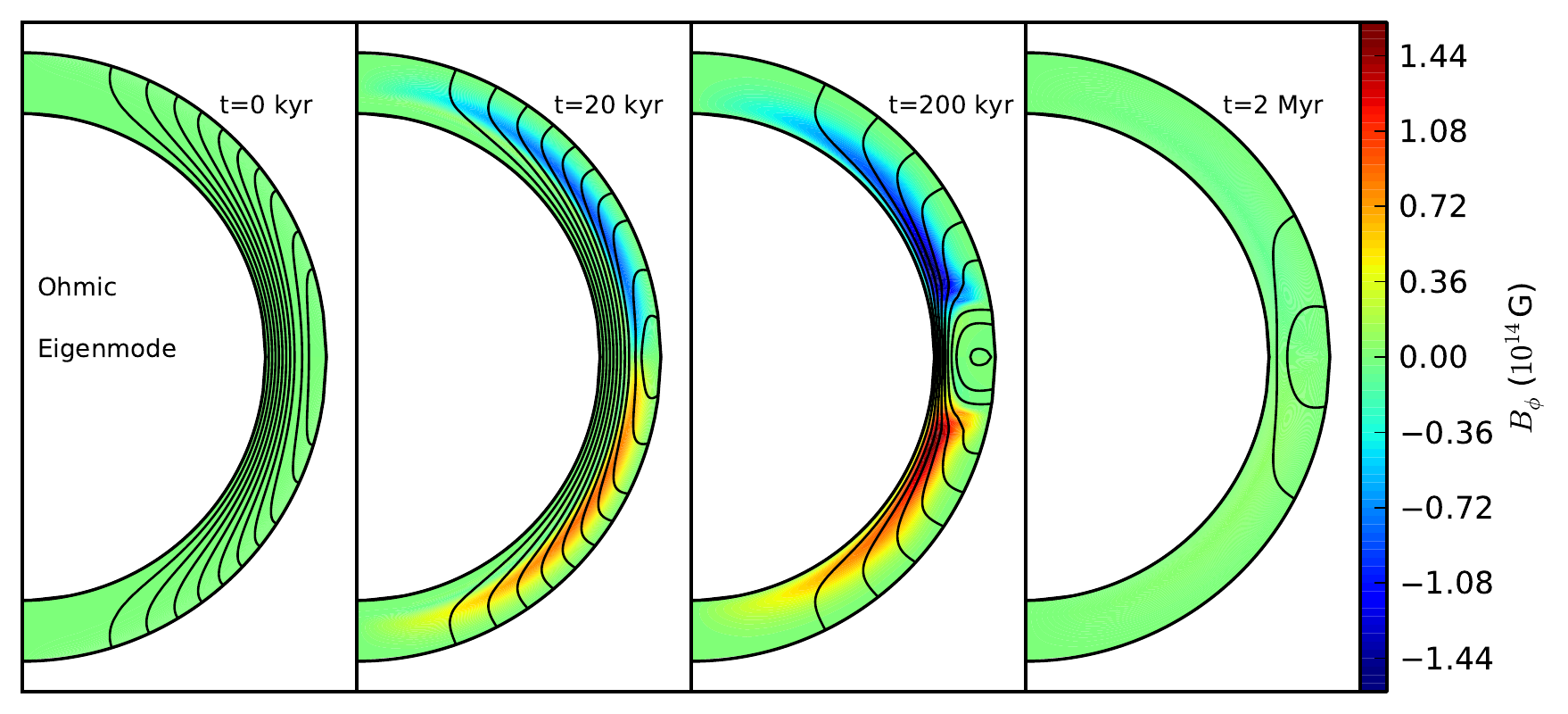}
\caption{Snapshots of the evolution of the magnetic field for the cases described in Section \ref{PPOLOIDAL}, at t=0 kyr, t=20 kyr, t=200 kyr and t=2 Myr. The black lines are ten equally spaced contours of constant $\Psi$ chosen in multiples of $\Psi_{max}/10$ where $\Psi_{max}$ is the maximum value of $\Psi$ at t=0 kyr for each model, the colour scale is the toroidal field. The thickness of the crust has been expanded by a factor of two, to allow clear depiction of the field. {\it Top Panel}. Barotropic equilibrium. At early time it develops an $\ell=2$ toroidal field with positive polarity in the northern hemisphere, which pushes the poloidal field lines closer to the pole, later the toroidal field breaks into multipoles and oscillates. Eventually Hall evolution saturates with the toroidal field consisting only of an $\ell=2$ toroidal field while the poloidal field decays preserving its structure. {\it Middle Panel}. Hall equilibrium. It takes much longer for this case to develop a toroidal field and start evolution. The toroidal field consists mainly of an $\ell=2$ field, and eventually Hall evolution saturates.   {\it Bottom Panel}. Ohmic eigenmode. The field initially develops a $\ell=2$ toroidal field with negative polarity in the northern hemisphere, which pushes the poloidal field lines to the equator. Subsequently the toroidal field forms higher multipoles, with eventual saturation of Hall evolution and relaxation to a state with a toroidal $\ell=2$ field and a poloidal field which preserves its structure. Note that the toroidal field developed is much stronger than the other two cases, as the differential rotation in the initial state is much greater than the other two cases, see Figure \ref{Fig:1}. }
\label{Fig:2}
\end{figure*}

\subsubsection{Barotropic Equilibrium}

\begin{figure}
\centering
\includegraphics[width=1.00\columnwidth]{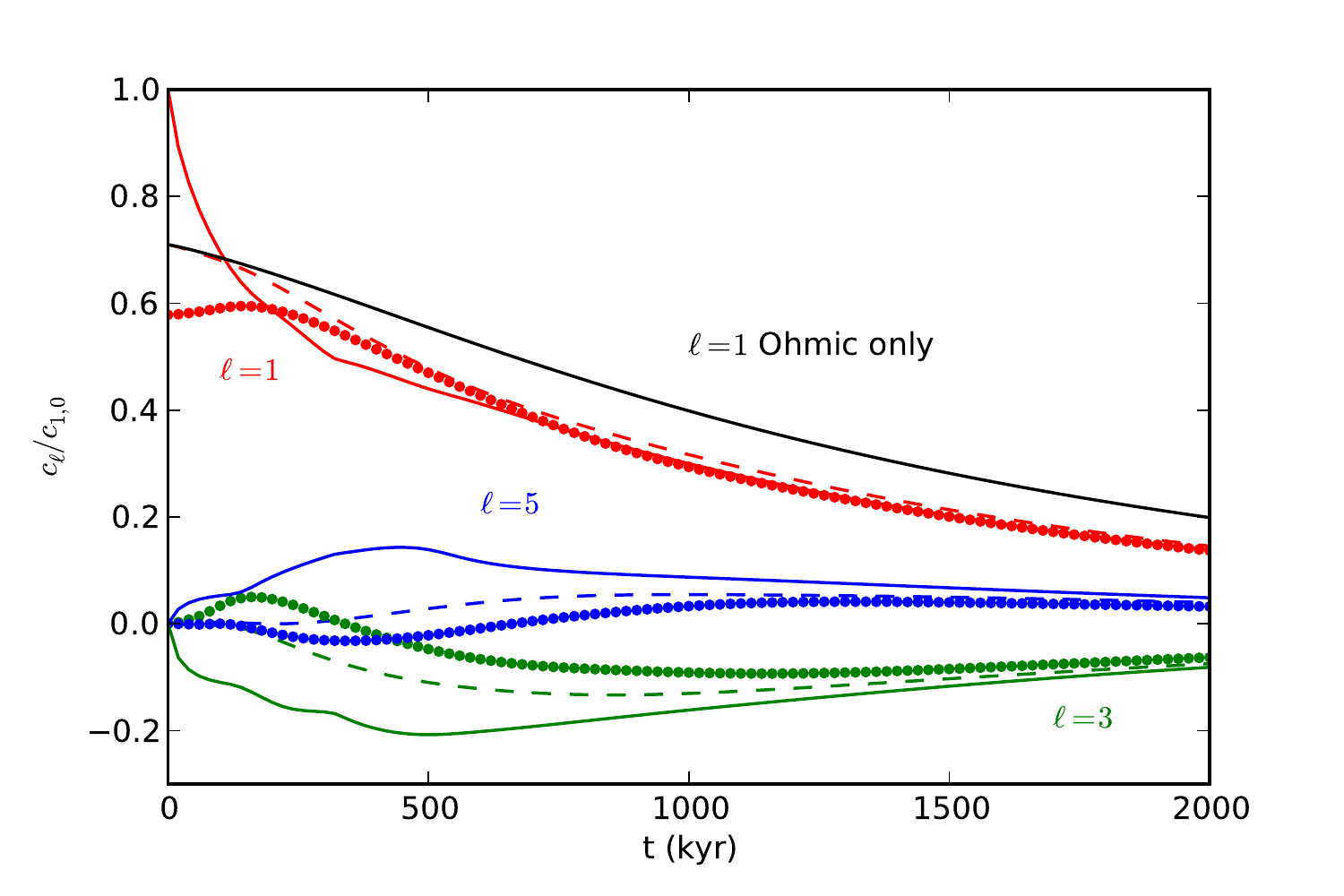}
\caption{Multipolar decomposition $c_\ell$, of the poloidal field on the surface of the neutron star for the cases discussed in Section \ref{PPOLOIDAL}. The red lines correspond to the amplitude of the $\ell=1$ component, the green to $\ell=3$ and the blue to $\ell=5$, while dotted correspond to barotropic MHD equilibrium, dashed to Hall equilibrium and solid to Ohmic eigenmode. The black solid line is the evolution of a field starting at Hall equilibrium under the influence of Ohmic dissipation only. The early evolution is clearly seen to depend on the initial conditions, with either increase or decrease of the $\ell=1$ component and a parallel behaviour of the $\ell=3$ component, with some oscillatory behaviour at intermediate times, while eventually  all three cases converge to the same structure the $\ell=1$ component being about twice as strong as the $\ell=3$. }
\label{Fig:3}
\end{figure}

Starting with the barotropic equilibrium (red curve in Fig.~\ref{Fig:1}), the fact that the electrons are moving more slowly further out leads to winding of the poloidal field lines that forms a toroidal quadrupole with positive polarity in the northern hemisphere. This is shown in the top panel of Figure \ref{Fig:2}. Subsequently, the poloidal currents associated with the newly-formed toroidal field wind the poloidal field lines, trying to align $\Psi$ with $I$ and pushing the poloidal field lines closer to the pole. This increases the dipole ($\ell=1$) and octupole component ($\ell=3$, we choose the polarity of a positive octupole being positive near the poles and negative at the equator) of the poloidal magnetic field emerging from the surface (Fig.~\ref{Fig:3}, dotted lines), despite the total magnetic energy decreasing because of Ohmic dissipation. 

\begin{figure}
\centering
\includegraphics[width=1.00\columnwidth]{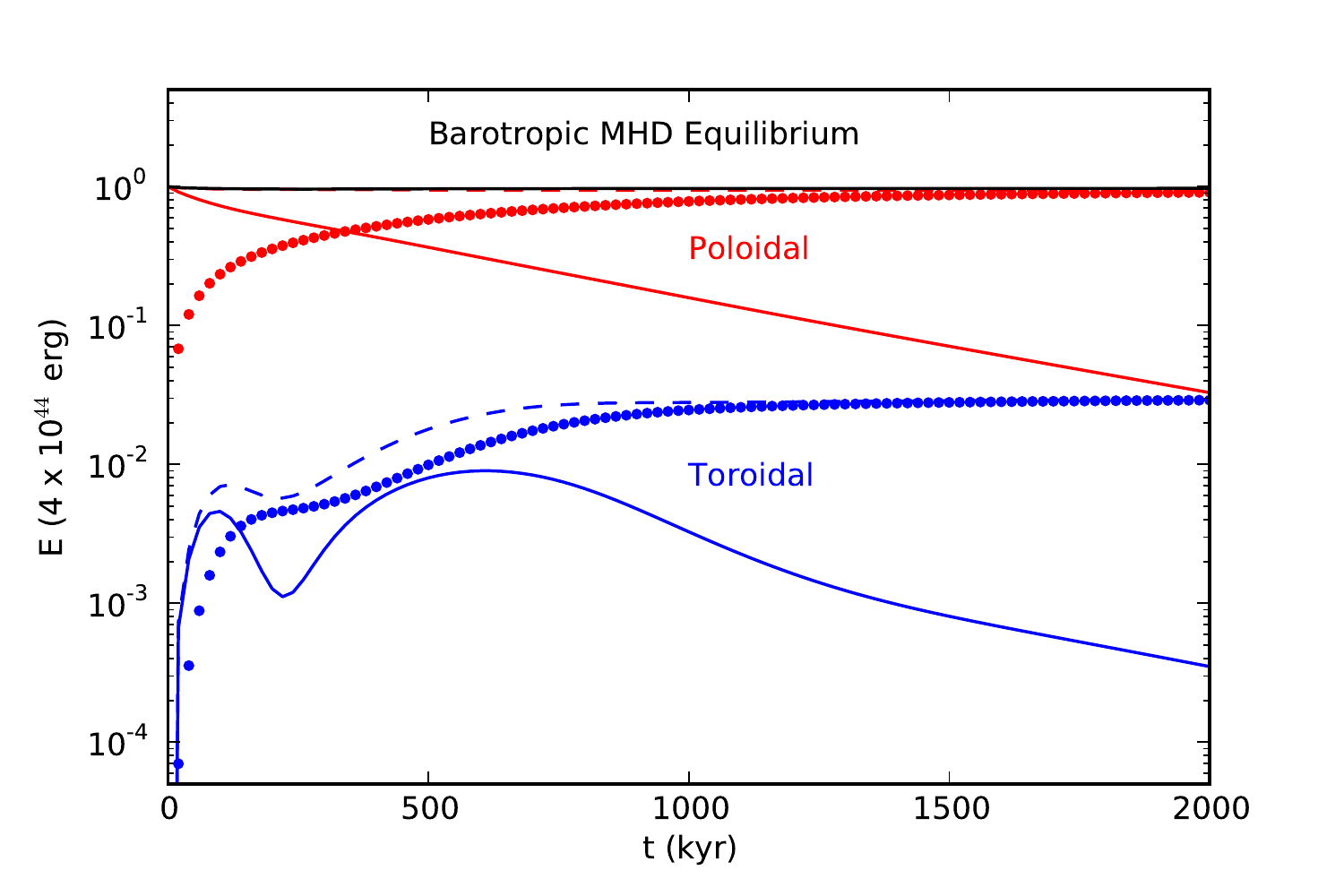}
\caption{The energy of the various components of the field and Ohmic losses during the evolution of a field starting at barotropic MHD equilibrium. The red lines correspond to the poloidal field, the blue to the toroidal field, the solid lines correspond to the energy of the field, the dotted line is the time integrated Ohmic losses through poloidal or toroidal field, the dashed is the sum of the the solid and the dotted line of the respective colour, the total energy is plotted in black to demonstrate that the code is conserving energy globally. Initially there is transfer of energy from the poloidal field to the toroidal, however between $100$ and $1000$ kyr there are also periods where the energy is transferred back to the poloidal field from the toroidal, shown at times where blue dashed line decreases. Eventually the field relaxes to a state where the toroidal field contains about $1\%$ of the total energy, while the poloidal and toroidal energies dissipate at the same rate. }
\label{Fig:4}
\end{figure}

This initial stage in which the $\ell=1$ and $\ell=3$ components grow lasts for about one Hall timescale, during which the toroidal field increases. After that, the intensity of the toroidal field decreases as part of its energy is returned to the poloidal field and part of it is Ohmically dissipated (Fig.~\ref{Fig:4}). The surviving toroidal field breaks into higher oscillating multipoles which form whistler waves propagating in the poloidal background. While this process is taking place, the surface $\ell=1$ component decreases, the surface $\ell=3$ changes polarity and reaches a maximum absolute value, later decreasing.

\begin{figure*}
\centering
\includegraphics[width=1.40\columnwidth]{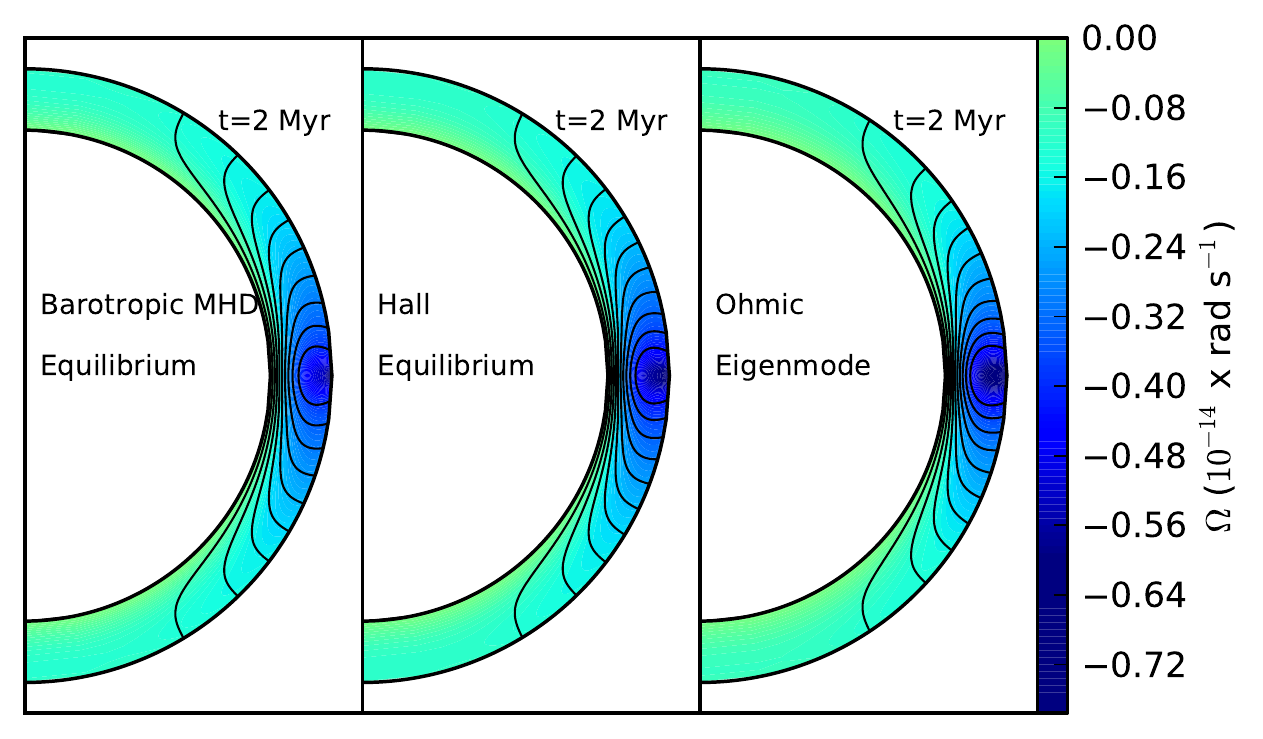}
\caption{Plot of the poloidal field lines and $\Omega$ at t=2 Myr of the three cases discussed in Section \ref{PPOLOIDAL}, showing that surfaces of constant $\Psi$ (poloidal field lines) coincide with surfaces of constant $\Omega$ (colour contours). The thickness of the crust has been expanded by a factor of two, to allow clear depiction of the field.}
\label{Fig:5}
\end{figure*}

Eventually, we find that after a few Hall timescales, the evolution saturates. The fields decay Ohmically, with an almost constant ratio of the different $\ell$ components. We can understand this saturation by looking at the profile of $\Omega$, shown in Figure \ref{Fig:5} (left panel). This shows clearly that the value of $\Omega$ is constant along each poloidal field line. Remarkably, the poloidal magnetic field relaxes to an isorotation state where $\Omega(\Psi)$ so that each poloidal magnetic field line rotates rigidly. In this state, the surface poloidal field consists mainly of an $\ell=1$ component and a $\ell=3$ component whose intensity about half of the $\ell=1$, while the toroidal field represents less than $1\%$ of the magnetic energy.


\begin{figure}
\centering
\includegraphics[width=1.00\columnwidth]{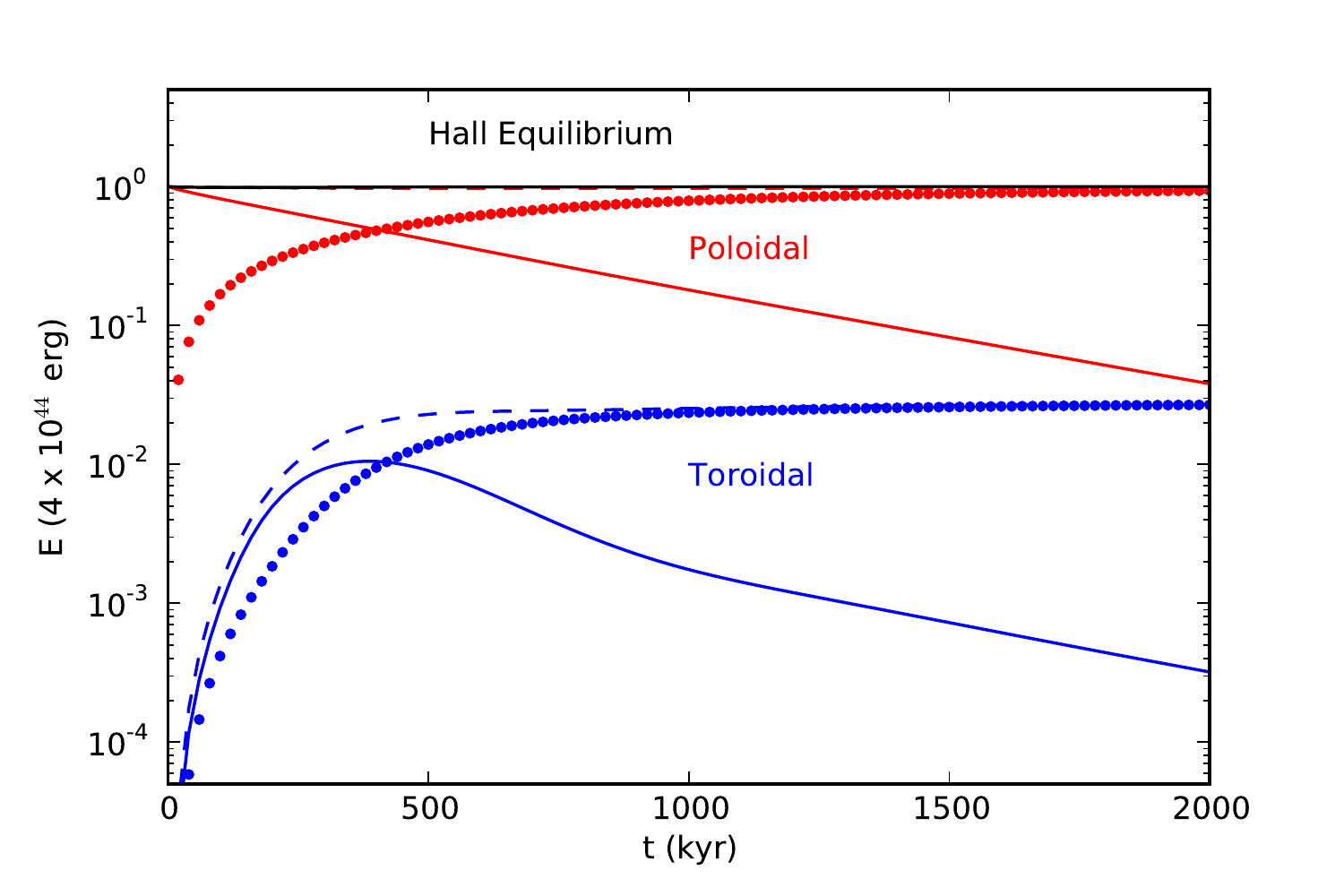}
\caption{The energy of the various components of the field and Ohmic losses during the evolution of a field starting at Hall equilibrium, the colour and line coding is the same as in Figure \ref{Fig:4}. As the field starts in the state of Hall equilibrium it takes much longer to develop a significant toroidal component with the subsequent energetic behaviour similar to that of Figure \ref{Fig:4}.}
\label{Fig:6}
\end{figure}
\begin{figure}
\centering
\includegraphics[width=1.00\columnwidth]{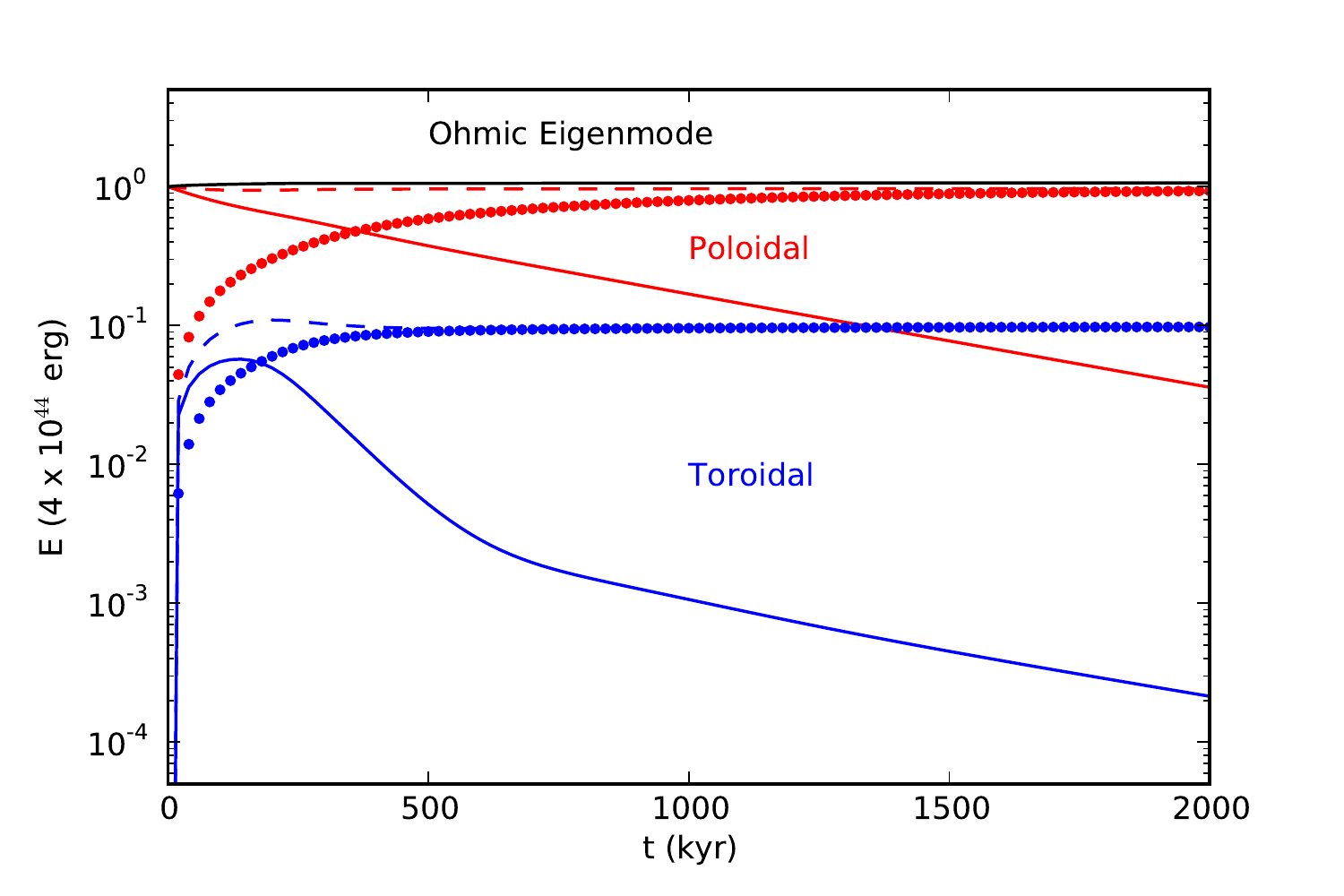}
\caption{The energy of the various components of the field and Ohmic losses during the evolution of a field starting at the fundamental Ohmic eigenmode, the colour and line coding is the same as in Figure \ref{Fig:4}. The toroidal field develops early in the evolution and reaches about $10\%$ of the total magnetic energy at around t=100 kyr, after that it decreases being transferred back to the poloidal field and lost Ohmically, as seen from the decrease in the blue dashed line . At later times the toroidal field corresponds to about $1\%$ of the total energy as in the other cases.}
\label{Fig:7}
\end{figure}

\subsubsection{Hall Equilibrium}

Next, we choose a dipole field in Hall equilibrium as an initial condition. The evolution is shown in the middle panel of Figure \ref{Fig:2}. Initially, the rigid rotation of the electrons means that there is no winding of the field. Instead, on a longer timescale Ohmic dissipation pushes the field slowly out of Hall equilibrium, so that the electrons do not rotate rigidly any longer. This differential rotation leads to winding of the poloidal field-lines into the toroidal  direction, and the appearance of a quadrupolar toroidal field of negative polarity in the northern hemisphere. Larger resistivity leads to faster generation of the toroidal field, confirming that the initial evolution is driven by Ohmic losses. Indeed, this Hall equilibrium can be expressed as a sum of the Ohmic eigenmodes. Because higher Ohmic modes dissipate faster the system is pushed towards the lower ones.

The toroidal field generated is of the opposite sign to the MHD equilibrium case, and acts on the poloidal field leading to a decrease in the $\ell=1$ component on the surface of the star and generating a negative $\ell=3$ component (dashed lines in Figure \ref{Fig:3}). The toroidal field remains in the $\ell=2$ state and decays because of energy losses back to the poloidal field and Ohmic dissipation. After a few Hall timescales, Hall evolution saturates, and we find again that the poloidal relaxes to an isorotation state with $\Omega(\Psi)$ (middle panel of Fig.~\ref{Fig:5}). The isorotation state again consists mainly of a $\ell=1$ and $\ell=3$ multipole components on the surface of the star with the $\ell=3$ having half the intensity of the $\ell=1$ and a toroidal field carrying less than $1\%$ of the magnetic field (Fig.~\ref{Fig:6}).

\subsubsection{Ohmic eigenmode}

Figure~\ref{Fig:2}, bottom panel, shows the evolution using an Ohmic eigenmode as initial condition. The initial winding of the poloidal field is such that the poloidal field lines generate a toroidal quadrupolar field with negative polarity in the northern hemisphere within a Hall timescale. This toroidal field winds-up the poloidal field lines trying to align $\Psi$ with $I$, decreasing the surface $\ell=1$ component and generating a negative $\ell=3$, Figure \ref{Fig:3}, while the poloidal field lines are pushed towards the equator. After this initial stage the toroidal field contains higher multipoles. After a few Hall timescales, similar to the other two cases discussed above, Hall evolution saturates with $\Omega(\Psi)$, and the surface decomposition of the poloidal field consisting mainly of $\ell=1$ and an $\ell=3$ component with about half the intensity of the latter and a toroidal field containing less than $1\%$ of the total magnetic energy (Fig.~\ref{Fig:7}).

The early evolution of the Ohmic eigenmode contrasts with the evolution of the barotropic MHD equilibrium. The different sign of the differential rotation leads to the opposite polarity toroidal field, and the opposite motion of the poloidal field lines. However, it is interesting to see that at later times the toroidal field has little memory of the initial conditions, which is wiped out through the whistler waves. Both cases eventually relax to the isorotation state.

\begin{figure*}
\centering
\includegraphics[width=1.60\columnwidth]{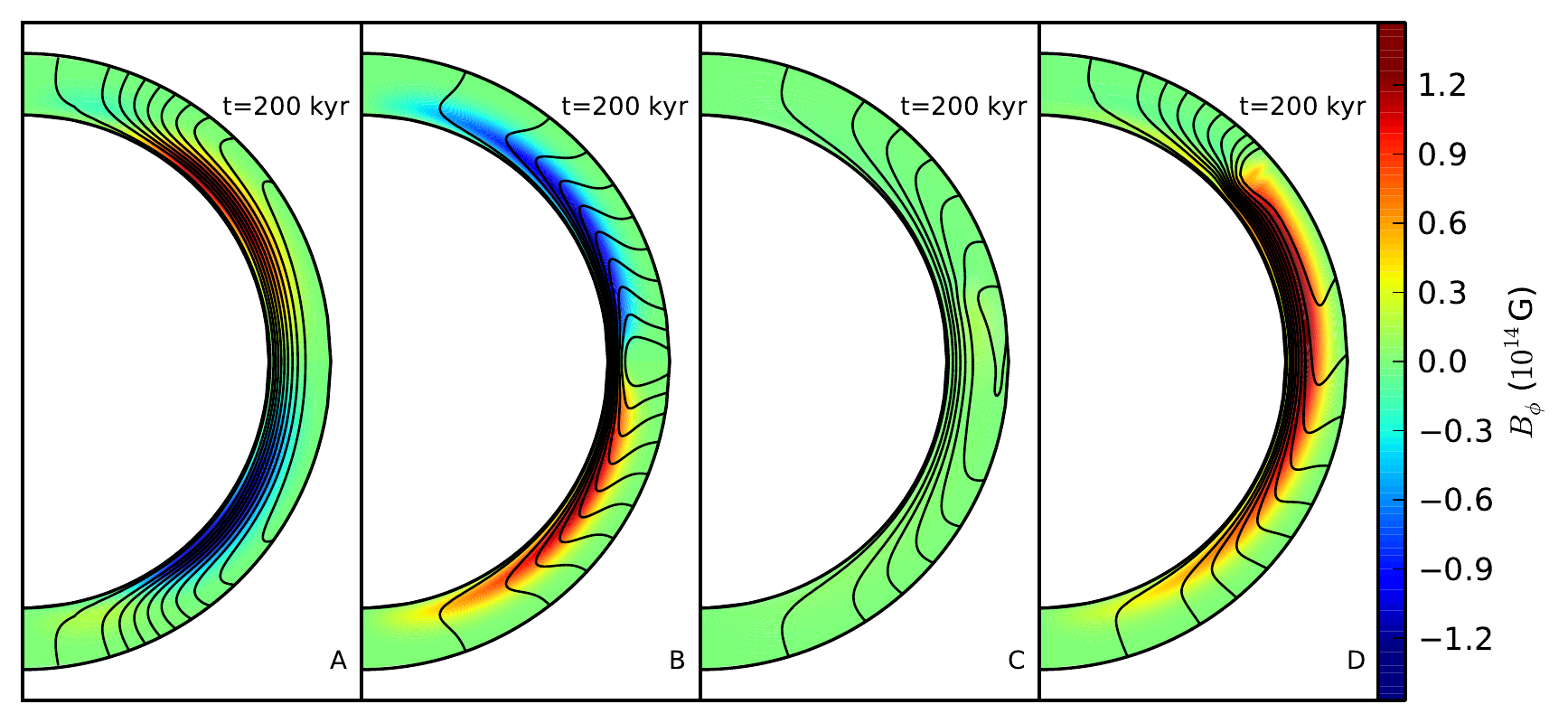}
\caption{A snapshot of the field structure at t=200 kyr of the four cases discussed in Section \ref{PTM}. The thickness of the crust has been expanded by a factor of two, to allow clear depiction of the field. {\it Plot A}, shows the field structure of a mixed $\ell=1$ poloidal with an $\ell=2$ toroidal with positive polarity in the northern hemisphere containing the same amount of energy, chosen as an extreme case of the early behaviour of the Barotropic MHD equilibrium field. The action of $I$ onto $\Psi$ pushes the field to the poles.  {\it Plot B}, shows the field structure of a mixed $\ell=1$ poloidal with an $\ell=2$ toroidal with negative polarity in the northern hemisphere containing the same amount energy, chosen as an extreme case of the early behaviour of the Ohmic eigenmode field. The action of $I$ onto $\Psi$ pushes the field to to equator.  {\it Plot C}, shows the field structure of a mixed $\ell=1$ poloidal a toroidal field confined in the closed poloidal field lines with positive polarity, the toroidal field contains about $0.1$ of the total energy. The toroidal field moves along surfaces of constant $\chi$ in the north direction dragging the poloidal field and leading to an asymmetric structures. {\it Plot D}, shows the field structure of a mixed $\ell=1$ poloidal with an $\ell=1$ toroidal with positive polarity, the fields contain the same amount of energy. Similarly to Plot C, the toroidal moves along surfaces of constant $\chi$ in the north direction dragging the poloidal field and leading to an asymmetric structures, but given the stronger toroidal field it has a stronger effect to the poloidal field.    }
\label{Fig:8}
\end{figure*}

\subsection{Poloidal and toroidal initial fields}
\label{PTM}

Following the results of the simulations with purely poloidal fields as initial conditions, we performed simulations with mixed poloidal and toroidal fields. Motivated by the fact that the polarity of the toroidal field generated determines the early evolution of the field, we performed simulations starting with mixed dipole poloidal field and toroidal $\ell=2$ with either polarity, with the poloidal and toroidal components containing the same amount of energy. In addition to those, we investigate two other families of initial conditions: a mixed poloidal dipole and toroidal $\ell=1$ containing the same amount of energy, and a poloidal dipole with a toroidal field confined in the closed field lines, where the toroidal field contains about $10\%$ of the energy. 

In the cases where we started with a mixed poloidal dipole and $\ell=2$ toroidal field, containing the same amount of energy, we observed the same behaviour as in our earlier simulations starting with poloidal-only barotropic equilibrium or Ohmic eigenmode initial conditions, but exaggerated. Figure \ref{Fig:8} panels A and B show the distortion of the poloidal field lines in each case. If the toroidal quadrupole is positive in the northern hemisphere, the early evolution qualitatively resembles that of the barotropic equilibrium. The $\ell=1$ component of the surface poloidal field increases by $\sim 20\%$ within a Hall timescale, accompanied by a significant formation of a positive $\ell=3$ poloidal component, while energy is lost from the toroidal field to the poloidal field and Ohmic dissipation. It takes only a few Hall timescales for the toroidal field to reduce its energy to $1\%$ of the total, while breaking into higher multipoles, while the poloidal field reaches an isorotation state. If the toroidal field has the opposite polarity, then the early evolution is similar to that of the Ohmic eigenmode, with the surface $\ell=1$ component decreasing by $\sim 30\%$ within a Hall time scale, and the development of a negative $\ell=3$. In the meantime, energy is lost from the toroidal field to poloidal, with the toroidal field eventually representing about $1\%$ of the total energy while breaking into higher multipoles. At later times the surface $\ell=1$ component increases slowly, while the $\ell=3$ evolves in a symmetric way. Eventually the poloidal field evolves to the isorotation state. 

The combination of a dipole poloidal field with a toroidal field which is either confined in the closed field lines, or is in the form a $\ell=1$ leads to a different early behaviour, as shown in panels C and D of Figure \ref{Fig:8}. In particular, the poloidal field evolves to be no longer symmetric about the equator. Depending on the polarity of the toroidal field, it moves either northwards or southwards because of the advection of $I$ along $\chi$ surfaces, and the poloidal field lines move also, trying to align themselves with surfaces of constant $I$. The later behaviour obeys the general pattern seen in our other simulations, that the toroidal field loses a significant amount of energy, and eventually represents $\sim 1\%$ of the total energy.

\section{Discussion}

In this section we discuss the implications of our results for the magnetic evolution of neutron stars. We outline the general properties of Hall and Ohmic evolution in the crust, discuss the observational signatures, and the constraints of such simulations.

\subsection{Hall Evolution}

In this work we have simulated a variety of initial conditions: purely toroidal fields, purely poloidal fields and mixed toroidal and poloidal fields. Despite the non-linearity of the equations, the evolution follows some basic steps that we outline below. 

A purely toroidal field does not generate any poloidal field, this is a consequence of axial symmetry, while it evolves with $I$ being advected along surfaces of constant $\chi$. Formally, a toroidal field is in equilibrium once surfaces of constant $I$ coincide with the surfaces of constant $\chi$, or $I=I(\chi)$. However, this state of equilibrium is impossible to be reached in a crust as surfaces of constant $\chi$ cross the crust-core boundary, while $I$ is confined in the crust. Thus, Hall evolution leads to the formation of strong currents, and eventually pushes the field to the crust-core boundary where it forms a shock and dissipates, at a Hall, rather than an Ohmic timescale. Once the intensity of the field drops, the ferocity of the Hall effect decreases and the dissipation continues on an Ohmic timescale. As $\chi$ also depends on the electron density, simulations with constant density give qualitatively opposite results compared to that of a stratified density crust, therefore such results should be approached with caution when applied to realistic neutron star crusts. 

Once a poloidal field is included the evolution becomes more complicated, yet more interesting for realistic neutron star applications. The general picture that emerges is as follows. The first couple of Hall time scales of the evolution is very sensitive to the initial conditions. Any state out of Hall equilibrium starts evolving immediately, the details of the evolution depend on the way the initial conditions deviate from that of a Hall equilibrium. If, for instance, the term arising from the misalignment of $\Omega$ with $\Psi$ is dominant, the field will generate a strong toroidal field which subsequently will act onto the poloidal field by changing the relative strength of the multipoles. If a significant toroidal field is present, then it will be its evolution because of motion of $I$ along $\chi$ surfaces that will be evident in early stages, while the poloidal field will become asymmetric because of the term that tries to align $\Psi$ with $I$. 

After this initial stage, the toroidal field may break into higher multipoles. Energetically, if the initial or the generated toroidal field contained more than a few percent of the total energy, its energy decreases, both by being Ohmically dissipated and by being transferred back to the poloidal field through the distortion of the poloidal field by the poloidal currents. In the meantime, the poloidal field evolves slowly towards isorotation, with $\Omega=\Omega(\Psi)$. In this state, the angular velocity of the electrons is constant along poloidal field lines, analagous to Ferraro's law in MHD. The field reaches this state after a few (5--10) Hall timescales. Eventually the field relaxes to a state where the poloidal field dominates, its decomposition on the surface of the star is mainly a dipole and an octupole component with the latter having about half the strength of the former. The toroidal field is much weaker, representing about $1\%$ of the total magnetic energy, and it has a quadrupole structure. We note that this saturation is apparent in previous studies, i.e. \cite{Hollerbach:2002}, after the end of the oscillatory behaviour, although they do not show results for the electron angular velocity profile. The number of the oscillations depends on their $R_{M}$, with higher Reynolds numbers allowing more oscillations before the saturation. 

\subsection{Observable properties}

The magnetic field is an element of key importance in the observable properties of neutron stars. In particular the structure of the field in the crust can influence the thermal conductivity, the intensity of the magnetic dipole emanating from the surface determines the spin-down behaviour and magnetar activity can be related to the Lorentz forces that can overcome the mechanical strength of the crust or the twist of the external field and lead to magnetospheric activity. The determination of observable properties requires a detailed study of the other physical properties of the neutron star that are not discussed in our Hall evolution study, while using the outcome of a Hall simulation as input. Nevertheless, there are some effects on observables that can be extracted from these simulations. 

The surface temperature is influenced by the magnetic field through two basic effects. First, thermal conductivity depends on the magnetic field, being stronger along magnetic field lines, and weaker perpendicular to them. Second, Ohmic dissipation converts magnetic energy into thermal energy. In this process there is a feedback loop, as also the electric conductivity depends on the temperature, complicating the problem. Heat transfer by phonons could also be important for very strong fields. In this work we can only address the first point, as for the latter a thermo-magnetic evolution scheme is essential \citep{Pons:2009}. The evolution of the models discussed in section \ref{POLOIDAL} suggests that in the case where the generated, or assumed, toroidal field has positive polarity in the northern hemisphere (barotropic equilibrium initial condition), one should expect a hotter pole and a much cooler equator, as the field lines emanating near the pole penetrate less deep the star, thus they guide more heat in a smaller surface. The contrary should hold in the case where the toroidal field is a quadrupole with negative polarity in the northern hemisphere (e.g.~Ohmic mode initial condition). In this case, less heat is deposited in a larger area around the poles, leading to a relatively cooler polar cap. These observables will be reflected in the X-ray pulse profile, however as pointed out by \cite{Vigano:2013}, the observed pulsed fraction is sensitive to the geometry and the absorption leading to degeneracies.  

Spin-down luminosity depends on the dipole component of the magnetic field, since this is the one surviving to the large distances of the light cylinder. In section \ref{POLOIDAL}, we concluded that for initial conditions of a barotropic MHD equilibrium, which is the most realistic case, the dipole component of the surface field increases in the early stage of the evolution, while the total magnetic energy is always decreasing. Such an increase of the dipole component is a natural way to explain braking indices clearly less than $3$, as it has been observed in all cases where this is possible \citep{Livingstone:2007, Espinoza:2011}. Further work is necessary to see whether the observed braking indices can be explained quantitatively in this picture. 
 
Magnetar activity is believed to occur through two basic mechanisms. It can either initiate internally, in regions where the magnetic field is so strong that it fractures the crust \citep{Thompson:1996}, or it can have its origin in the magnetosphere, with the field being twisted because of motion of the footpoints on the neutron star, up to a point that becomes unstable and releases the energy explosively \citep{Lyutikov:2006}. The first case has been studied by \cite{Perna:2011} who found that younger neutron stars with strong magnetic fields are likely to exhibit Soft Gamma-ray Repeater (SGR) behaviour, while the older ones are most likely to be seen as Anomalous X-ray Pulsars (AXPs). The latter case has been discussed by \cite{Parfrey:2012}, who simulated the evolution of a twisted magnetosphere, assuming that a region of field lines on the neutron star is displaced, and they found that the magnetospheric field can be twisted up to a critical value, after which currents become that strong leading to a burst. In our calculations we find that even if we assume a strong toroidal field initially it only takes a few Hall timescales for the poloidal field to dominate. As the twisted magnetosphere model implies a strong toroidal field, it is likely that this scenario is applicable to the early evolution, while older ones can exhibit magnetar activity through fractures, as even without strong toroidal field there can be a significant $\bm{j} \times \bm{B}$ term.

\subsection{Boundary conditions}

In our simulation we have chosen the Meissner boundary condition for the crust-core interface and the vacuum boundary condition for the surface of the star. The Meissner condition, is applicable to a type-I superconductor which cannot be penetrated by the magnetic field. The sharp decrease in the number of free electrons and the abundance of neutrons in the outer core, suggest that the evolution of the magnetic field in the core will be strongly dominated by ambipolar diffusion in that region. The timescales for the evolution due to this effect are different for the solenoidal and the irrotational components of the fluid velocity and magnetic force, with the solenoidal diffusing  faster by several orders of magnitude. On the  other hand, the Hall timescale in general is shorter than that of ambipolar diffusion. In this case, even if the field penetrates the crust-core interface, the evolution in the crust has no effect in the core, and the inner boundary condition can be treated as a slowly evolving field, or even a constant field. In the case where the two timescales are comparable, a realistic simulation should take into account the feedback of the crust to the core. In this case the inclusion of a realistic inner boundary condition may affect the compression of the field near the crust-core boundary, as this is the result of demanding all the magnetic flux to be concentrated in the crust. In the opposite case, if the Hall time scale is short enough we do not expect the basic conclusion related to the evolution of a toroidal field to change, as even in this case the Hall effect pushes the magnetic field in the crust core interface while it takes much longer for it to move in the core because of the much longer timescale, thus it will form strong currents in this region which shall lead to fast dissipation. 

With respect to the outer boundary condition we have assumed that the neutron star is surrounded by a vacuum, which is appropriate for a non-rotating neutron star. A rotating neutron star will be surrounded by at least a Goldreich-Julian plasma \citep{GoldreichJulian:1969}, while there have been observations which suggest that the actual plasma density is a few orders of magnitude larger than the Goldreich-Julian one \citep{Lyutikov:2004, Rafikov:2005}. Therefore, even for slow rotation, currents can flow in the magnetosphere and support toroidal fields. In this case, the condition of $I=0$ on the surface of the neutron star can be relaxed, and instead of assuming a vacuum multipole for the field outside the star, a force-free solution should be used. Note that since the rotation of magnetars is in general slow, the ratio of the light cylinder to the radius of the neutron star is a large number, thus the force-free solution does not need to take into account the relativistic terms. Apart from the fact that such a boundary condition is more realistic, it could also address the issue of magnetar activity through twisted magnetospheres. 

We have also applied our outer boundary condition within the star, at densities $\approx 10^{11}\ {\rm g\ cm^{-3}}$, assuming that the lower density regions have a vacuum (potential) field. As we argue in \S 2, a strong field will overpower the mechanical strength of the outer regions of the star. This leaves interesting questions about how the crust responds to the magnetic deformations in this region.

\section{Conclusions}

In this paper we have discussed the evolution of the magnetic field of a neutron star under the influence of the Hall effect and we have explained the key behaviours in the light of analytic arguments. We have confirmed that the evolution of a toroidal field is essentially advection along surfaces of constant $\chi$. The evolution of a mixed poloidal and toroidal field follows a three stage regime (Figures \ref{Fig:2} and \ref{Fig:3}): The early evolution is the response of the field to any unbalanced terms of the equation of Hall evolution, a dipole poloidal field with differential rotation of the electron fluid along the field lines bends and generates a quadrupole toroidal field, a poloidal field with an $\ell=1$ toroidal field out of equilibrium will start its evolution with $I$ moving along surfaces of constant $\chi$ and dragging the poloidal field with it, this stage lasts a couple of Hall timescales. We remark that if we choose a barotropic MHD equilibrium as the initial state we observed an increase of the surface dipole component of the magnetic field at the early stages of evolution. The intermediate stage follows for a few Hall timescales depending on the initial $R_{M}$, during this stage the toroidal field breaks into multipoles which oscillate as whistler waves. Through this process the poloidal field relaxes to an isorotation with $\Omega=\Omega(\Psi)$, (Figure \ref{Fig:5}) while a weaker toroidal field survives in the form of a quadrupole containing about $1\%$ of the total energy. Intermediate age neutron stars that have undergone Hall evolution, should end up with having a poloidal field that consists mainly of an dipole and an octupole component.

\section*{Acknowledgements}

We thank David Tsang for insightful discussions and his invaluable assistance with the plots. We thank Vicky Kaspi, Andreas Reisenegger, Maxim Lyutikov, Jonathan Braithwaite, Crist\'obal Armaza and Dong Lai for inspiring discussions. KNG is supported by the Centre de Recherche en Astrophysique du Qu\'ebec.  AC is supported by an NSERC Discovery Grant and is an Associate Member of the CIFAR Cosmology and Gravity program.

\bibliographystyle{mnras}
\bibliography{BibTex.bib}

\section*{Appendix}

We integrate forward equations \ref{dPSI} and \ref{dI} using a finite difference code. We define a discrete grid $i,j$ with $i_{0}$, $i_{c}$ and $i_{s}$ the points corresponding to the centre of the star, the crust-core interface and the surface of the star, and $j_{1}$, $j_{2}$ the points corresponding to the south and the north pole respectively, and $j=0$ as the equator. The correspondence of the coordinates is such, so that $r=i~\delta r$ and $\cos\theta=\mu=j \delta \mu$, where $\delta r= 1/(i_{s}-i_{0})$ and $\delta \mu = 2/(j_{2}-j_{1})$. At each time step $n$ we calculate the spatial derivatives for $\Psi$ and $I$ using the values of the previous time step $n-1$. Then we substitute in the right-hand-side of equations \ref{dPSI} and \ref{dI} and we multiply by $\delta t$ to find the values at $t+\delta t$. To calculate the spatial derivatives of $I$ and $\Psi$ we use the following formulae:
\begin{eqnarray}
\Psi_{r} = \frac{\Psi_{i+1, j}-\Psi_{i-1,j}}{2\delta r} \,,
\end{eqnarray}
\begin{eqnarray}
\Psi_{rr}=\frac{\Psi_{i-1, j}-2\Psi_{i,j}+\Psi_{i+1,j}}{\delta r^{2}}\,, 
\end{eqnarray}
\begin{eqnarray}
\Psi_{rrr}=\frac{\Psi_{i+2, j} - 2\Psi_{i+1, j} +2 \Psi_{i-1, j} -\Psi_{i-2, j}}{2\delta r^{3}}\,.
\end{eqnarray}
\begin{eqnarray}
\Psi_{rr\mu}=\frac{\Psi_{rr~i, j+1}-\Psi_{rr~i,j-1}}{2\delta \mu}
\end{eqnarray}
The same scheme is used for the differentiation with respect to $\mu$, and for the derivatives of $I$. For the field outside the star we fit a multipole poloidal field, and we use two rows of dummy points outside the star to evaluate the derivatives at $i=i_{s}$ and $i=i_{s}-1$. We assume a row of dummy points in the $j$ boundaries of the star, so that  $\Psi_{i,j_{1}-1}=-\Psi_{i,j_{1}+1}$ and  $\Psi_{i,j_{2}+1}=-\Psi_{i,j_{2}-1}$ to evaluate the derivatives $\Psi_{\mu \mu \mu}$ at $j=j_1+1$ and $j=j_2-1$. At the crust-core boundary we assume that $\Psi_{i_{c}-1, j}=-\Psi_{i_{c}+1, j}$ to evaluate the $\Psi_{rrr}$ derivative at $i=i_{c}+1$. We use the Meissner boundary condition at the base of the crust by setting $\Psi$ equal to zero, meaning that the magnetic field does not thread the core. We set the toroidal field equal to zero both at the inner and outer surface by requiring $I=0$. At each points with $j_{1}<j<j_{2}$ and $i_{s}<i\leq i_{c}$ we evaluate the spatial derivatives using the above scheme, and we evolve the values by $\delta t=0.25 h \delta r \delta \mu$ where $h$ is a Courant parameter which is determined by the largest electron velocity in the grid, that shrinks the time step to avoid numerical instabilities. The values of the conductivity, density and their derivatives that were used in the code were calculated a priori and stored, as they do not evolve with time. 

We have tested our code in cases where analytical solutions are known. We have started with a magnetic configuration and we let it decay Ohmically confirming that each eigenmode indeed decays at the right timescale. The detailed evolution of the toroidal field is also in agreement with previous simulations and the analytical arguments. The code conserves the total energy $E(t)$ locally and globally, when the energy of the magnetic field, Ohmic losses and any energy exchanged with the outside of the star are included, with an accuracy better than $(E(t)-E(0))/E(0)<0.05$ in the simulations presented in this paper. Simulations with toroidal fields only have an even better energy conservation performance. The simulations presented in this paper use a 100$\times$100 grid for the variation of $r$ and $\mu$ in the crust. We have run test cases with double the resolution and we found the same behaviour. We have run simulations with a  crust being $0.2r_{*}$ and we have found that the magnetic field evolves in a similar way with the time evolution to be about 4 time slower. 

In the numerical code we used the following normalisation for equation (\ref{dB}):
\begin{eqnarray}
1.6\times 10^{6} \frac{\partial \bm{B}_{14}}{\partial t_{\rm yr} } = - \tilde{\nabla} \times  \left(\frac{\tilde{\nabla} \times \bm{B}_{14}}{n_{\rm e,0}} \times \bm{B}_{14}\right) \nonumber\\
-0.02\tilde{\nabla} \times \left(\frac{\tilde{\nabla} \times \bm{B}_{14}}{\sigma_{0}}\right)\,,
\end{eqnarray}
where $B_{14}=B/10^{14}$G, $t_{yr}=t/3.15\times 10^{7}$s, $\tilde{\nabla}$ is the del operator where all lengths are normalized to $10^{6}$cm, $n_{\rm e, 0}=n_{\rm e}/2.5\times 10^{34}$cm$^{-3}$, $\sigma_{0}=\sigma/(1.8\times 10^{23})$s$^{-1}$, in the simulations performed the maximum initial value of $R_{M}$ is in the range of 50 to 100. 

Below we give initial conditions for $\Psi$ and $I$ for the simulations presented in this paper using the above normalization, where $r$ is given in units of $10^{6}$ cm:\\
i) Figure \ref{Fig:2}, top panel: $I=0$, $\Psi = (-61573.524/r+2.9591204\times10^6r^2-5.8418851\times10^7r^4+2.1719352\times10^8r^5-4.0045241\times10^8r^6+4.4664864\times10^8r^7-3.1630100\times10^8r^8+1.3996384\times10^8r^9-3.5479564\times10^7r^{10}+3.9482757\times10^6r^{11})(1-\mu^{2})$. This field corresponds to a barotropic equilibrium field with $S_{B}=$const.~ and $\rho \propto n_{\rm e}^{7/4}$ which allows $Y_{\rm e}$ to vary by a factor of 30 from the base to the surface of the crust.\\
ii) Figure \ref{Fig:2}, middle panel: $I=0$,   $\Psi_{H}=(9747.235035r^2-531.0200763/r-80179.06283r^4+1.703400527\times 10^5r^5-1.570329631\times 10^5 r^6+70059.32069r^7-12403.39223r^8)(1-\mu^{2})$. This field is the Hall equilibrium for rigid electron rotation, for the choice of $n_{\rm e}$ we have made, and is used for the poloidal field in the mixed poloidal and toroidal initial conditions.\\  
iii) Figure \ref{Fig:2}, bottom panel: $I=0$,   $\Psi=((0.011408524(-16.385\cos(16.385r)r+\sin(16.385r)))/r-(0.0091699536(\cos(16.385r)+16.385\sin(16.385r)r))/r) (1-\mu^{2})$.\\  
iv) Figure \ref{Fig:8} A: $\Psi=2^{-1/2}\Psi_{H}$, $I=7.6871\times 10^{6}(0.9-r)^2(1-r)^2 \mu(1-\mu^2)$.\\ 
v) Figure \ref{Fig:8} B, $\Psi=2^{-1/2}\Psi_{H}$, $I=-7.6871\times 10^{6}(0.9-r)^2(1-r)^2 \mu(1-\mu^2)$.\\
vi) Figure \ref{Fig:8} C, $\Psi=10^{-1/2}\Psi_{H}$, $I=0$ for $\Psi<\Psi_{0}$, $I=2665(\Psi-\Psi_{0})$ for $\Psi>\Psi_{0}$, where $\Psi_{0}$ is the value of $\Psi$ on the equator. \\
vii) Figure \ref{Fig:8} D,  $\Psi=2^{-1/2}\Psi_{H}$, $I=3.437\times 10^{5}(0.9-r)^2(1-r)^2(1-\mu^2)$. \\

\end{document}